
\magnification=\magstep1
\vsize=23truecm
\hsize=16truecm
\baselineskip=.6cm
\hoffset=0.truecm
\voffset=0.truecm
\font\ti=cmbx10 scaled\magstep2

\vglue 2cm

\centerline{\ti A quantum group version of}
\centerline{\ti  quantum gauge theories in two dimensions}

\vskip 2cm
\centerline{M. Karowski
\footnote{*}{Supported by DFG, SFB 288
``Differentialgeometrie und Quantenphysik"}
\footnote{$^1$}{e-mail: karowski@vax1.physik.fu-berlin.dbp.de}}
\centerline{R. Schrader
\footnote{$^2$}{e-mail: schrader@vax1.physik.fu-berlin.dbp.de}}

\vskip .5cm
\centerline{Institut f\"ur Theoretische Physik}
\centerline{Freie Universit\"at Berlin, Germany}

\vskip 2cm

\noindent{\bf Abstract:}

\medskip
For the special case of the quantum group $SL_q (2,{\bf C})\
 (q= \exp \pi i/r,\  r\ge 3)$ we present an alternative approach
 to quantum gauge theories in two dimensions.
We exhibit the similarities to Witten's combinatorial approach which
 is based on ideas of Migdal.
The main ingredient is the Turaev-Viro combinatorial construction of
 topological invariants of closed, compact 3-manifolds and its
 extension to arbitrary compact 3-manifolds as given by the authors in
 collaboration with W. M\"uller.
\vfill\eject

\beginsection 1. The classical group construction of Yang Mills
 theories in two dimensions.

Based on ideas originally invented by Migdal [M] for the case $\Sigma
 ={\bf R}^2$, Witten [W] gave a combinatorial formulation of quantum
 gauge theories on an arbitrary compact 2-dimensional Riemann manifold
 $(\Sigma, \eta)$ with metric $\eta$.
(For a path integral approach, see also [F1], [F2]).
It is part of a program devoted to a calculation of the volume of the moduli
space of flat connections on compact 2-manifolds.

For  the readers convenience and for later comparison we recall the
 construction in [W] for the case that $\Sigma$ is oriented.
Let $X$ denote a cell decomposition of $\Sigma$ (actually a
 triangulation would suffice) with oriented i-cells
 $c^i\ (0\le i\le 2)$.
As in standard lattice gauge theories a discrete version of a parallel
 transport is a map $U(\cdot)$ from the set of oriented 1-cells $c^1$
 into $G$ such that $U(c_1^*)=U(c_1)^{-1}$,
where $c_1^*$ equals $c_1$ but with the opposite  orientation.
The gauge group is the set of all maps $V(\cdot)$ from the set of zero
 cells $c^0$ into $G,\ c^0\mapsto V(c^0)$.
This gauge group operates on the parallel transporters via $V(\cdot) :
 U(\cdot) \mapsto U_V(\cdot)$ with
$$U_V(c^1)=V(c_i^0) U(c^1)V(c^0_f)^{-1} \eqno(1)$$
where $c^0_i$ and $c_f^0$ are the initial and final vertices of $c^1$
 respectively.
For a given configuration $U(\cdot)$ and an oriented 2-cell $c^2$ let
$$U(c^2) = \prod_{c^1\in \partial c^2} U(c^1) \eqno(2)$$
be the parallel transporter around $c^2$, where the product is taken
 in cyclic order, counterclockwise say.
The conjugacy class of
$U(c^2)$ is unique and gauge invariant.
For the given metric $\eta$ on $\Sigma$ let $\rho(c^2)$ be the area of
 $c^2$.
Also let $\chi_\alpha$ denote a complete set of irreducible characters
 of $G$ and $c_2(\alpha)$ the value of the Casimir operator in the
 representation $\alpha$ defined by $\chi_\alpha$.
With dim $\alpha =\chi_\alpha(e)$ being the dimension of this
 representation, set
$$\Gamma (U(c^2),\rho(c^2))=\sum_\alpha \dim \alpha \ \chi_\alpha
 (U(c^2))\cdot \exp\big(-e^2\rho(c^2)c_2(\alpha)/2\big)\eqno(3)$$
Here $e\not= 0$ is a real parameter, the coupling constant.
Now define
$$Z_X (e^2,\eta) = \int \prod_{c^1\in X} dU(c^1) \prod_{c^2\in X}
 \Gamma (U(c^2),\rho(c^2)),\eqno(4)$$
where $dU$ is the normalized Haar measure on $G$.
The main result of Witten in this context is that $Z_X (e^2,\eta)$ is
 independent of the particular choice of $X$.
Thus if $X$ is a triangulation, an elementary argument shows for
 example invariance under Alexander moves [A] and this suffices to
 establish the claim.
Therefore $Z_X(e^2,\eta)$ gives rise to an invariant denoted by
 $Z_\Sigma (e^2\rho)$ in [W], where $\rho$ is the total area of
 $\Sigma$ for given $\eta$.
By the argument given in [W], $Z_\Sigma(0)$ can be viewed as the
volume of the moduli space of flat $G$-connections on $\Sigma$.
Furthermore, if $\Sigma= \Sigma^g$ is a Riemann surface of genus $g$,
 then
$$Z_{\Sigma^g} (e^2\rho) = \sum_{\alpha}  {\exp\big(-e^2\rho
 c_2(\alpha)/2\big)\over (\dim \alpha)^{2 g-2} } . \eqno (5)$$
Now $Z_\Sigma(e^2\rho)$ was generalized as follows.
Suppose $\Sigma=\Sigma_n$ has $n$ holes i.e. the boundary
 $\partial\Sigma$ of $\Sigma$ consists of components
$\partial\Sigma_i\cong S^1\ (i=1,\dots,n)$. For the induced cell
decomposition $\partial X_i$ of $\partial \Sigma_i$ and given
$U(\cdot)$ let $U(\partial X_i)$ be the parallel
transport around $\partial X_i$ and set
$$Z_{\Sigma_n} (e^2 \rho; \alpha_1,\dots,\alpha_n) =
 \int \prod_{c^1\in X} dU(c^1) \prod_{c^2\in X}
 \Gamma (U(c^2),\rho(c^2))
\prod_i\overline{\chi_{\alpha_i}\big(U(\partial X_i)\big)}
.\eqno(6)$$
For the particular case where $\Sigma=\Sigma_n^g$ has genus $g$,
this gives
$$Z_{\Sigma_n^g} (e^2 \rho; \alpha_1,\dots,\alpha_n) =
 \delta_{\alpha_1,\dots,\alpha_n} {\exp -e^2\rho c_2(\alpha)/2\over
 (\dim \alpha_1)^{2g-2+n} }. \eqno(7)$$
Here $\delta_{\alpha_1,\dots ,\alpha_n} =1$ if all $\alpha_i$ are
 equal and zero otherwise.

Suppose now $\Sigma$ with area $\rho$ is cut into pieces $\Sigma_\mu$
with areas $\rho_\mu\ (\rho=\sum_\mu\rho_\mu)$ by cutting along
circles $C_i\ (i=1,\dots,\mu)$ labelled by representation $\alpha_i$
and resulting in a colouring $\underline\alpha_\mu$ on the boundary
circles of $\Sigma_\mu$. Then one has the surgery formula
$$Z_\Sigma (e^2\rho) = \sum_{\underline\alpha}\prod_\mu
 Z_{\Sigma_\mu}(e^2\rho, \underline\alpha_\mu)
.\eqno(8)$$

By cutting $\Sigma^g$ into $(2g-2)$ spheres, each with 3 holes, in the
 standard way, relations (7) and (8) directly lead to (5).

 \beginsection  2. The quantum group formulation.

The aim of this article is to present a quantum group version for the
 case $SL_q(2,{\bf C}) \break (q=\exp i\pi/r,\  r\ge 3)$.
This has the advantage that one may directly work at $e^2=0$ such that
 the resulting partition function is metric independent.
Note that in (3) the infinite sum is well defined  due to the
 presence of the cut-off  $\exp -e^2\rho(c^2) c_2(\alpha)/2$.
In the $SL_q(2,{\bf R})$ case the parameter $r$ acts as a regularizing
 factor, since we will only have to deal with representations $\alpha$
 labelled as elements in $ {\cal I} = \{ 0,{1\over 2},\dots, {r\over
 2} -1\}$.
In the classical limit $r\to \infty$ these results agree with those
obtained by Witten for the case $e^2=0$ and $G=SU(2)$.

The idea is based on the combinatorial construction of Turaev and Viro
 [TV] of topological invariants of closed compact 3-manifolds using
 $SL_q(2,{\bf C})$ and its extension to arbitrary compact 3-manifolds
 as given in [KMS].

In fact, given this Turaev-Viro state sum $Z_{TV}(M)$ for compact
 3-manifolds $M$, we define for any compact 2-manifold $\Sigma$
the following quantities  which are complex numbers
$$Z(\Sigma) =Z_{TV} (\Sigma\times I) \eqno(9)$$
and where $I$ is the unit interval.
Note that $\Sigma$ need not be orientable.

We want to show that this simple construction relating
2-dimensional theories to 3-dimensional ones has interesting
features with properties closely related to the ones given in
the preceeding section.
In fact in [KS]  in case the compact 3 manifold $M$ is oriented we
 introduced partition functions $Z(M,G_{\underline \alpha})$ for
 coloured graphs $G_{\underline \alpha}$ on the boundary $\partial M$.
Roughly speaking a coloured graph is a graph $G$ on $\partial M$,
 where to each maximal interval $\ell_i$ of $G$ one associates  an
 element $\alpha_i$ of ${\cal I}$.
These partition functions are homotopy invariants of the coloured graphs
 and are useful in order to obtain surgery relations in analogy to
 (8).
Given this construction, we may extend (9) in case $\Sigma$ is
 oriented to define
$$Z(\Sigma, G_{\underline\alpha}^\ell, G_{\underline \beta}^r)= Z_{TV}
 (\Sigma \times I, G_{\underline\alpha}^\ell \cup
 G_{\underline\beta}^r)\eqno(10)$$
where $G_{\underline\alpha}^r$ is a coloured graph on $\Sigma\times
 \{0\}$ and $G_{\underline\beta}^\ell$ is another coloured graph on
 $\Sigma \times \{1\}$.
We are interested in the special case where again $\Sigma =\Sigma_n^g$
 is a Riemann surface of genus $g$ with $n$ holes.
Let $G_{\underline\beta}^r$ be the empty graph and let
 $G_{\underline\alpha}^\ell = G_{\underline\alpha}\ (\underline\alpha
 =(\alpha_1,\dots, \alpha_n))$ be the coloured graph consisting of $n$
 circles $S^1$ around these $n$ holes and carrying the colours
 $\alpha_1,\dots,\alpha_n$ respectively.

Then by the methods and results in [KS] (see in particular Lemmas 5.1
 and 5.3 and appendix B) one has in analogy to (7)
$$Z(\Sigma_n^g,G_{\underline\alpha}) =
 \delta_{\alpha_1,\dots,\alpha_n} {1\over (w^2_{\alpha_1})^{2g-2+n}
 },\eqno(11)$$
where
$$w_\alpha^2 = (2\alpha+1)_{-q} = (-1)^{2\alpha} {q^{2\alpha+1}
 -q^{-2\alpha-1}\over q-q^{-1} }  \eqno(12)$$
is the $q$-dimension associated to the representation  $\alpha$.
Obviously $w_\alpha^2$ replaces $\dim\alpha=2\alpha+1$.
In fact $\displaystyle{\lim_{r\to \infty}} w^2_\alpha =
  (-1)^{2\alpha}(2\alpha+1)$.

With otherwise the same notation as in (8) it is also easy to prove
 the same surgery formula
$$Z(\Sigma)= \sum_{\underline\alpha}\prod_\mu Z(\Sigma_\mu,
 G_{\underline\alpha_\mu})  .\eqno(13)$$
In particular we obtain
$$Z(\Sigma^g) = \sum_\alpha {1\over (w_\alpha^2)^{2g-2}}\eqno (14)$$
which compares with (5) and which has a classical limit for $g\ge 2$
equal to $\zeta(2g-2)$.

We note that the right hand side is essentially the Verlinde formula
 [V].
In fact, the fusion matrices in the present context are given by
$$N^k_{ij} = \cases{ 1 &if $k\le i+j,\ j\le i+k,\ i\le k+j,\
   r-2\ge i+j+k\in {\bf Z}$\cr 0& otherwise\cr}$$
with $i,j,k \in  {\cal I} $,  such that $N_{ij}^k = N_{ji}^k
 =N_{kj}^i$.
Also all these matrixes $N^k$ commute and with
$$\vec N^2 = \sum_k (N^k)^2 \eqno(15)$$
we may rewrite (14) in the form (see e.g. [KS], appendix A and C for
 the present context)
$$Z(\Sigma^g) = w^{-2g+2} {\rm trace}\ (\vec N^2)^{2g-2} \eqno(16)$$
with $w^2 =\sum_\alpha w_\alpha^4$.

This compares with the partition function for the Chern-Simons theory
 for the group $G=SU(2)$ at level $k=r-2$
$$Z_{c.s.} (\Sigma^g\times S^1) = {\rm trace}\ (\vec N^2)^{2g-2}.
 \eqno(17)$$
\bigskip\bigskip\noindent
{\bf Acknowledgments:} The above discussion was initiated by a
stimulating conversation with M. Thaddeus, whom we would like
to thank.

\beginsection References:

\item{[A]} Alexander, J.W. : The combinatorial theory of complexes,
 Ann. Math. {\bf 31}, 294-322 (1930).

\item{[F1]} Fine, D. : Quantum Yang Mills theory on the two sphere,
 Commun. Math. Phys. {\bf 134}, 273-292 (1990).\par
\item{[F2]} Fine, D. : Quantum Yang Mills on a Riemann surface,
 Commun. Math. Phys. {\bf 140}, 321-338 (1991).\par
\item{[KMS]} Karowski, M. , M\"uller, W. , Schrader, R. : State sum
 invariants of compact 3-manifolds with boundary and 6j-symbols, to
 appear in Journ. Phys. A.\par
\item{[KS]} Karowski, M. , Schrader, R. : A combinatorial approach to
 topological quantum field theories and invariants of graphs,
 submitted for publication in Commun. Math. Phys.\par
\item{[M]} Migdal, A. : Zh. Eksp. Teor. Fig. {\bf 69}, 810 (1970)
 (Sov. Phys. Jetp.{\bf 42}, 413).\par
\item{[TV]} Turaev, V.G. , Viro, O.Y. : State sum of 3-manifolds and
 quantum 6j-symbols, to appear in Topology.\par
\item{[V]} Verlinde, E. : Fusion rules and modular transformations in
 2d conformal field theory, Nucl. Phys. {\bf B300}, 360-376
 (1988).\par
\item{[W]} Witten, E. : On quantum gauge theories in two dimensions,
 Commun. Math. Phys. {\bf 141}, 153-209 (1991).\par

\bye